\documentclass[12pt]{article}

\newcommand{\Title}[1]{\begin{center}{\large\bf #1}\vskip 4pt\end{center}}
\newcommand{\Author}[1]{\sl #1}
\newcommand{\Email}[1]{\footnote{Email: {\tt #1}}}
\newcommand{\Affil}[1]{\begin{center}#1\end{center}}
\newcommand{\Abstract}[1]{\begin{abstract}#1\end{abstract}}

\begin{document}

\Title{The Wheeler -- DeWitt Quantum Geometrodynamics:\\
its fundamental problems and tendencies of their resolution}

\begin{center}
\Author{T.P. Shestakova\Email{shestakova@phys.rsu.ru}}
\end{center}
\Affil{Department of Theoretical and Computational Physics,
Southern Federal University\footnote{former Rostov State University},
Sorge St. 5, Rostov-on-Don 344090, Russia}

\Abstract{The paper is devoted to fundamental problems of the Wheeler --
DeWitt quantum geometrodynamics, which was the first attempt to
apply quantum principles to the Universe as a whole. Our purpose is
to find out the origin of these problems and follow up their
consequences. We start from Dirac generalized Hamiltonian dynamics
as a cornerstone on which the Wheeler -- DeWitt theory is based. We
remind the main statements of the famous DeWitt's paper of 1967 and
discuss the flaws of the theory: the well-known problem of time,
the problem of Hilbert space and others. In the concluding part of
the paper we consider new tendencies and approaches to quantum
geometrodynamics appeared in the last decade.}

{\bf I. Introduction.} There is no doubt that the first significant
attempt to construct full quantum theory of gravity was presented
in the paper by DeWitt of 1967 \cite{DeWitt}. As DeWitt mentioned,
as soon as quantum theory had been invented, attempts to apply it
to gravitational field had been made, among others, by Rosenfeld
and Bergmann (see \cite{DeWitt} for references), who had faced
enormous obstacles. The obstacles consisted in nonlinear properties
of the Einstein theory that made all calculations very tedious, but
the main difficulty was that the nature of general relativity as a
completely covariant theory ran counter to efforts to build a
Hamiltonian formulation of it as the first step on the way of its
quantization. The difficulty was referred to as ``the problem of
constraints''. Meanwhile, in 1950s Dirac published his outline of a
general Hamiltonian theory \cite{Dirac1,Dirac2} which was in
principle applicable to any system with constraints, in particular,
to gravitational field. The next important step was done by
Arnowitt, Deser and Misner \cite{ADM} who proposed a special
parametrization of gravitational variables that made the
construction of Hamiltonian formalism easier and admitted a clear
interpretation. The third source of DeWitt theory was the ideas of
Wheeler concerning a wave functional describing a state of
gravitational field \cite{Wheeler1,Wheeler2}.

The Wheeler -- DeWitt quantum geometrodynamics encountered a number
of fundamental problems which cannot be resolved in its own limits.
The purpose of this paper is to find out their origin and follow up
their consequences. So, we start from Dirac generalized Hamiltonian
dynamics as a cornerstone on which the Wheeler -- DeWitt theory is
based. We shall see how Dirac postulates determined the structure
of quantum geometrodynamics and try to answer the question, if
application of the Dirac method to gravitational field was
justified. In the next part of the paper we shall consider new
tendencies and approaches to quantum geometrodynamics appeared in
the last decade and show that the new approaches suggest a revision
of the Wheeler -- DeWitt theory.

{\bf II. Dirac approach to quantization of constrained systems.}
``The problem of constraints'' implies that field equations are not
independent, some of them involve no second time derivatives,
i.e. they are not dynamical equations but constraints. The
objective of Dirac was to take into account the presence of
constraints in Hamiltonian formalism, while it was impossible to
do making use of a Hamiltonian constructed according a usual rule
$H=p_a\dot q_a-L$. It is well-known that Dirac proposed to replace
the ``original'' Hamiltonian $H$ by a new one adding to $H$ a
linear combination of constraints:
\begin{equation}
\label{tot.Ham}
H_T=H+c_m\varphi_m.
\end{equation}
Strictly speaking, this Dirac proposal looks like a postulate. It
may be interesting that in his paper of 1950 \cite{Dirac1}, Dirac
tried to ground it by means of some speculations based on a
variational procedure, but in his paper of 1958 \cite{Dirac2} and
his lectures \cite{Dirac3} he omitted these speculations just
introducing a {\it total} Hamiltonian (\ref{tot.Ham}).

It is worth noting that the generalized Hamiltonian dynamics is not
completely equivalent to Lagrangian formulation of the original
theory. In the Hamiltonian formalism the constraints generate
transformations of phase space variables, however, the group of
these transformations does not have to be equivalent to the group
of gauge transformations of Lagrangian theory. In this sense, the
task to construct a Hamiltonian formulation for any system
describable by an action functional has not been entirely solved
since instead of two equivalent formulations one obtains two
theories dealing with two different groups of transformations.

In quantization procedure the Hamiltonian (\ref{tot.Ham}) does not
take a significant place, and so does not a Schr\"odinger equation
with a Hamiltonian operator corresponding to the classical
expression (\ref{tot.Ham}). The central role is given, again, to
the constraints: each ``weak'' equation (by Dirac terminology)
$\varphi_m(q,p)=0$ after quantization becomes a condition on a
state vector, or wave functional, $\Psi$:
\begin{equation}
\label{Dir.cond}
\varphi_m\Psi=0.
\end{equation}
This is another postulate of Dirac, which cannot be justified by
the reference to the correspondence principle. Dirac immediately
noticed \cite{Dirac1} a new obstacle that is known as the ordering
problem. Let us emphasize that in quantum geometrodynamics the
latter gives rise to the problem of parametrization noninvariance,
when various forms of gravitational constraints, being equivalent
on the classical level, lead to different equations for a wave
functional, which are already non-reducible to each other.

One could raise a question: what is the significance of the Dirac
generalized dynamics for the development of quantum field theory?
One should confess that the success of quantum electrodynamics and
other gauge theories, which have advanced our understanding of
microworld, is due to other ideas and methods rather then the Dirac
generalized dynamics. The only theory that is essentially grounded
on the latter is the Wheeler -- DeWitt geometrodynamics, which,
however, has not made any verifiable predictions.

{\bf III. The ADM parametrization.} In the seminal paper \cite{ADM}
was introduced the following presentation for components of metric
tensor:
\begin{equation}
\label{ADM-metric}
ds^2=-(N^2-N_iN^j)dt^2+2N_idx^idt+g_{ij}dx^idx^j.
\end{equation}
Here, the lapse function, $N$, determines the interval of proper
time between two subsequent spacelike hypersurfaces, and the shift
functions, $N_i$, define the shift of a point under transition
from one hypersurface to another. The ADM parametrization introduces
in 4-dimensional spacetime a set of 3-dimensional hypersurfaces
(the so-called (3+1)-splitting). It enables one to write the
gravitational action in terms of the second fundamental form, or
extrinsic curvature tensor, $K_{ij}$, and intrinsic curvature
tensor, $^{(3)}R_{ij}$, and, eventually, after introducing
canonical momenta, in the Hamiltonian form:
\begin{eqnarray}
S&=&\int\!d^4x\sqrt{-g}R
 =\int\!d^4xN\sqrt{^{(3)}\!g}\,(K_{ij}K^{ij}-K^2+\,^{(3)}\!R)\nonumber\\
\label{ADM-action}
&=&\int\!d^4x(\pi\dot N+\pi^i\dot N_i+\pi^{ij}\dot g_{ij}
-N{\cal H}-N_i{\cal H}^i).
\end{eqnarray}
${\cal H}$ and ${\cal H}^i$ are the Hamiltonian and momentum
constraints,
\begin{equation}
\label{Ham.constr}
{\cal H}=\sqrt{^{(3)}\!g}\,(K_{ij}K^{ij}-K^2-\,^{(3)}\!R)
=G_{ijkl}\pi^{ij}\pi^{kl}-\sqrt{^{(3)}\!g}\;^{(3)}\!R;
\end{equation}
\begin{equation}
\label{Gijkl}
G_{ijkl}=\frac12\sqrt{^{(3)}\!g}\,(g_{ik}g_{jl}+g_{il}g_{jk}-g_{ij}g_{kl}).
\end{equation}
\begin{equation}
\label{mom.constr}
{\cal H}^i=-2\pi^{ij}_{;j}
=-2\partial _j\pi^{ij}-g^{il}(2\partial_k g_{jl}-\partial_lg_{jk})\pi^{jk}.
\end{equation}
The status of the two constraints is different: The momentum
constraints generate diffeomorphisms of 3-metric $g_{ij}$ and are
similar to constraints in the Yang -- Mills theory. But a basic
role is given to the Hamiltonian constraint, whose dynamical
character results from non-standard quadratic dependence of
${\cal H}$ from the momenta $\pi^{ij}$, and so the Hamiltonian
constraint has no analogy in other gauge theories. There is no
constraint in the theory which would generate transformations for
$N$, $N_i$.

In spite of some advantages of the ADM parametrization, we should
remember that it is just one of possible parametrizations that
allows us to construct Hamiltonian formalism for gravitation. To
give an example, let us refer to the work by Faddeev
\cite{Faddeev}, where the author introduces quite intricate
variables $q^{ij}=h^{0i}h^{0j}-h^{00}h^{ij}$,
$\lambda^0=\displaystyle\frac1{h^{00}}+1$,
$\lambda^i=\displaystyle\frac{h^{0i}}{h^{00}}$, where
$h^{\mu\nu}=\sqrt{-g}g^{\mu\nu}$, and Hamiltonian formalism is
constructed in terms of $q^{ij}$ and their momenta.

{\bf IV. The essence of the Wheeler -- DeWitt theory.} What are the
constraints to become in quantum theory? As DeWitt emphasized, they
cannot become operator equations, for otherwise the Hamiltonian
determined in (\ref{ADM-action}) would yield no dynamics at all.
Indeed, the gravitational Hamiltonian, which is just a linear
combination of the constraints, would become a zero operator. Then,
in accordance with the Dirac approach, the primary constraints
$\pi=0$, $\pi^i=0$ and secondary constraints ${\cal H}=0$, ${\cal
H}^i=0$ become conditions on the state vector:
\begin{equation}
\label{all-constr}
\pi\Psi=0,\quad
\pi^i\Psi=0,\quad
{\cal H}\Psi=0,\quad
{\cal H}^i\Psi=0.
\end{equation}

After the replacement of the momenta by functional differential
operators, $\pi=-i\displaystyle\frac{\delta}{\delta N}$,
$\pi^i=-i\displaystyle\frac{\delta}{\delta N_i}$, the first two
equations in (\ref{all-constr}) mean that the state vector does not
depend on $N$, $N_i$. The last, or momentum constraints are
interpreted as the conditions that a wave functional is invariant
under coordinate transformations of 3-metric. In common these
conditions lead to the conclusion that the wave function depends
only on 3-geometry. In this respect DeWitt followed to the idea by
Wheeler that the wave functional must be determined on the
superspace of all possible 3-geometries \cite{Wheeler1,Wheeler2}.
But the statement that the wave function depends only on 3-geometry
remains to be pure declarative: it has no mathematical realization.
As a matter of fact, the state vector always depends on a concrete
form of the metric.

There remains the third equation, ${\cal H}\Psi=0$, the famous
Wheeler -- DeWitt equation. Its solution corresponding to
observable physical Universe is supposed to be singled out by
appropriate boundary conditions. Again, since the Hamiltonian is a
linear combination of the constraints, one comes to the conclusion
that quantum geometrodynamics can never yield anything but a
static picture of the world:
\begin{equation}
\label{stat.pic}
H\Psi=\int\!d^{3}x(N{\cal H}+N_i{\cal H}^i)\Psi=0,\quad
H\Psi=i\frac{\partial\Psi}{\partial t},\quad
\frac{\partial\Psi}{\partial t}=0,
\end{equation}
and the Schr\"odinger equation loses its significance in quantum
geometrodynamics. DeWitt \cite{DeWitt} commented it as following:
Physical significance can be ascribed only to intrinsic dynamics of
the Universe. In the case of a finite world one has no preferable
physically relevant coordinates for the description of the
intrinsic dynamics, and ``the constraints are everything''. DeWitt
wrote that one of his task was to convince the reader that nothing
else but the constraints is needed. Has DeWitt really convinced us
that the constraints are everything that we need to understand
the quantum Universe?

{\bf V. Fundamental problems of the Wheeler -- DeWitt theory.} The
conclusion about the static picture of the world, or {\it the
problem of time}, is the most known problem of the Wheeler --
DeWitt theory. It creates other fundamental problems, some of them
we shall consider here briefly (for discussion, see
\cite{Isham,SS1,SS2}).

Imposing of constraints restricts the spectrum of the Hamiltonian
by the only value $E=0$ that creates {\it the problem of Hilbert
space}. The structure of Hilbert space is believed to be specified
if the inner product of state vectors is defined. Without a
well-defined inner product one cannot calculate averages of
physical quantities that raises doubts in the ability of the theory
to make predictions. The inner product is to conserve in time, so
some definition of time is required. In some approaches, time is
identified with a function of variables of configurational or phase
space. But in this case the status of time variable differs from
what it is in ordinary quantum mechanics, namely, an extrinsic
parameter related to an observer and marking changes in a physical
system.

Another problem which is closely connected with the problem of
time is {\it the problem of observables}. According to the Dirac
scheme, observables are quantities which have vanishing Poisson
brackets with constraints. It is indeed the true for
electrodynamics where all observables are gauge-invariant. But in
case of gravity this criterion leads to the conclusion that all
observables should not depend on time. Then one loses a
possibility to describe time evolution of a gravitational system
in terms of observables.

We have already mentioned above {\it the problem of
reparametrization noninvariance} which is inseparable from {\it the
ordering problem}: At the classical level the gravitational
constraints can be written in various equivalent forms while at the
quantum level, after replacing the momenta by operators, these
different forms of the constraints become nonequivalent. It is a
consequence of the fact that the supermetric in the configurational
space of all 3-metrics $g_{ij}$, which coincides with the inverse
of the DeWitt supermetric $G_{ijkl}$ (\ref{Gijkl}) under the choice
$N=1$ accepted in \cite{DeWitt}, depends, in general, on the lapse
function $N$. Hawking and Page were the first who explicitly
pointed to this dependence \cite{HP}. Then, the Wheeler -- DeWitt
equation ${\cal H}\Psi=0$ before solving the ordering problem can
be written as
\begin{equation}
\label{WDW-eq}
\left(G_{ijkl}(N)\frac{\delta}{\delta g_{ij}}
 \frac{\delta}{\delta g_{kl}}+\sqrt{^{(3)}\!g}\;^{(3)}\!R\right)\Psi=0.
\end{equation}
To choose any ordering one should impose an additional relation
between $N$ and $g_{ij}$ (it may be, in particular, the same
condition $N=1$ which implies that $N$ does not depend on
$g_{ij}$). It is important to understand that the ordering problem
cannot be solved without making use, explicitly or implicitly, of
the additional condition on $N$. Parametrization and this condition
together determine an ultimate form of the Wheeler -- DeWitt
equation. Further, the choice of another parametrization than
(\ref{ADM-metric}) would change the structure of the supermetric in
the configurational space of gravitational variables and the
structure of the differential operator in (\ref{WDW-eq}). It would
result in an analog of the Wheeler -- DeWitt equation, which, in
general, could not be reduced to the first one. Accordingly, the
two equations would have different solutions.

As has already said above, the ADM parametrization fixes
(3+1)-splitting of spacetime that enables one to distinguish
between spacelike and timelike geometrical object and to present
the action in the Hamiltonian form. However, one can apply this
procedure only if spacetime has the topology $R\times\Sigma$, where
$\Sigma$ is some 3-manifold. In any other case it is impossible to
introduce globally (in the whole spacetime) a set of spacelike
hypersurfaces without intersections and other singularities, and it
is impossible to introduce a global time. This is {\it the problem
of global structure of spacetime}.

{\bf VI. The criticism of the Wheeler -- DeWitt theory and the
problem of gauge invariance.} We can see that in the case of
gravity the application of the Dirac postulate, that constraints
become condition on a state vector, laconically expressed in
DeWitt's words ``the constraints are everything'', is the origin of
fundamental difficulties of the theory. So, was the application of
the Dirac method to gravitational field justified? Why the central
role in the theory was given to the constraints?

The difficulties of the Wheeler -- DeWitt theory have made a way
for its strong criticism. So, Isham \cite{Isham} doubted that there
is a real justification for extending the Dirac approach to
constraints that are quadratic functions of the momenta. He wrote:
``...although it may be heretical to suggest it, the Wheeler -­
DeWitt equation -- elegant though it be -- may be completely the
wrong way of formulating a quantum theory of gravity''.

At the classical level, the constraints is known to express gauge
invariance of the theory. It was initially believed that imposing
constraints at the quantum level would also ensure gauge invariance
of wave functional. Strictly speaking, the founders of quantum
geometrodynamics have not investigated this issue and its gauge
invariance has not been proved. It leads us to the next fundamental
problem: Could we consider quantum geometrodynamics as a
gauge-invariant theory?

It was pointed out in the work by Mercury and Montani \cite{MM1}
that, making use of the ADM formalism and fixing (3+1)-splitting,
one also chooses particular values for the lapse function $N$ and
the shift vector $N_i$. Therefore, this is equivalent to fix a
reference frame, so that gauge invariance breaks down and the
Hamiltonian constraint loses its sense and, with the latter, so
does the whole procedure of quantization. As the same authors wrote
in another paper \cite{MM2}, the ADM splitting is equivalent to a
kind of ``gauge fixing''.

Let us remind in this connection that the canonical quantization
based on the Dirac generalized dynamics is not the only way to
construct quantum theory. Another possibility is to appeal to the
path integral approach, which, in some respects, is more powerful.
There have been made several attempts to give a rigorous derivation
of the Wheeler -- DeWitt equation from a path integral (see, for
example, \cite{HH,BP}). The most accurate and consequent derivation
was made by Halliwell \cite{Hall}. However, in \cite{BP,Hall} the
so-called asymptotic boundary condition were used which are typical
for systems with asymptotic states explored in ordinary quantum
field theory, when particles in initial and final states are far
away from interaction region. In the case of gravitational field
one can speak about asymptotic states only if space is
asymptotically flat.

The problem of gauge invariance of quantum geometrodynamics was
thoroughly analysed in \cite{SSV1,SSV2}. It was argued that,
generally speaking, a universe with non-trivial topology does not
possess asymptotic states, and one cannot impose asymptotic
boundary conditions in this case. The equation for a wave function
of the Universe is proved to be gauge-dependent.

On the other hand, it was demonstrated that parametrization
noninvariance of the theory and its gauge noninvariance are
intimately related. Of course, parametrization noninvariance {\it
is not the same} that gauge noninvariance. Nevertheless, as was
shown above, to determine the form of the Wheeler -- DeWitt
equation, one has to impose an additional condition on $N$, which,
in a fact, is a gauge condition on $N$. (In the full theory, to
define the full Hamiltonian, one has to fix also conditions on the
shift vector $N_i$.) As was emphasized in \cite{Sh},
parametrization and gauge conditions together fix a reference
frame. To summarize, the belief that imposing the constraints on a
state vector (\ref{all-constr}) in quantum geometrodynamics ensures
its gauge invariance is just a misunderstanding of how the things
work.

{\bf VII. New tendencies: Evolutionary Quantum Geometrodynamics.}
As we have seen, the Wheeler -- DeWitt quantum geometrodynamics
failed to be proved to be a gauge invariant theory and the
constraints lose their significance. The analysis of its
fundamental problem leads to the conclusion that the Wheeler --
DeWitt theory needs to be revised.

On a deeper lever, the reason of the failure of this theory
consists in a complicated structure of general relativity itself.
Though the latter resembles in many aspects other gauge theories,
this resemblance should not be misleading. In particular, in the
theory of gravity gauge degrees of freedom cannot be considered as
somehow redundant: they determine the spacetime geometry and
contribute to gauge-invariant expressions like, for example,
4-curvature $R$. It has been argued in the recent paper
\cite{Leclerc}, that gravitational field is not entirely
specified by the constraints and dynamical equations in contrast
to electromagnetic field. It would be fair to say that we are
still far from ultimate understanding of all features of general
relativity even at the classical level.

Now return to a revision of the Wheeler -- DeWitt theory. It is
obvious enough that any solution of the problems of time and the
problem of Hilbert space requires the rejection of the Hamiltonian
constraint as a condition on the state vector or, at least, some
modification of the constraint. In the latter case the Wheeler --
DeWitt equation $H\Psi=0$ reduces to a stationary
Schr\"odinger-like equation $\tilde H\Psi=E\Psi$. A similar
modification was discussed yet by Weinberg \cite{Wein} and Unruh
\cite{Unruh} and aimed to solve the cosmological constant problem.
The ideas by Weinberg and Unruh were reproduced recently in
\cite{DF}, where gauge noninvariance of quantum cosmology is used
to introduce a particular gauge in which the cosmological
constraint is quantized. In \cite{DF} and other papers just minor
amendments to the Wheeler -- DeWitt theory were suggested which can
be considered as a remedy for a solution of some particular
problem, without a careful analysis of the principles on which a
consistent quantum geometrodynamics must be based.

Another way is to reject the Wheeler -- DeWitt equation at all and
to reestablish the role which Schr\"odinger equation plays in any
quantum theory and which it lost in quantum geometrodynamics. The
main tendency of the last decade consists in the replacement of the
static picture of the world of the Wheeler -- DeWitt theory by
evolutionary quantum gravity. This way is more fundamental since it
revises the foundations of the theory. It has been realized in
recent years that it is impossible to obtain the evolutionary
picture of the Universe without fixing a reference frame. In this
trend we should refer to the work by Brown and Kucha\v r \cite{BK}
where a privileged reference frame was fixed by introducing an
incoherent dust which plays the rule of ``a standard of space and
time''. In already mentioned works by Montani and his collaborators
\cite{MM1,MM2} (see also \cite{BM}) the so-called kinematical
action was introduced describing a dust reference fluid. It leads
to what the authors call ``Schr\"odinger quantum gravity''.

In the ``extended phase space'' approach to quantum
geometrodynamics \cite{SSV1,SSV2} a reference frame is introduced
by means of a gauge-fixing term in the action as it is used to be
done in quantum field theory. It was shown in \cite{SSV2} that the
gauge-fixing term admits the following interpretation: it
describes a subsystem of the Universe, some medium, whose equation
of state is determined by gauge conditions. The central part is
given to the Schr\"odinger equation for a wave function of the
Universe, the latter being derived from a path integral by
rigorous mathematical procedure.

A general feature of the above approaches is that they describe the
Universe evolution in a certain reference frame. In classical
general relativity we can, knowing a solution of Einstein equations
in one reference system, calculate what phenomena would take place
in another reference system. In quantum geometrodynamics we do not
know what reference frame would be preferable, how the solutions
corresponding to different reference frames are related to each
other and how they should be interpreted. As it often happens, the
developments of this field of study creates new problems and
puzzles.

\end{document}